\documentclass[twocolumn,english,aps,prl,reprint,superscriptaddress,floatfix]{revtex4-1}

\usepackage[LGR,T1]{fontenc}
\usepackage[latin9]{inputenc}
\usepackage{verbatim}
\usepackage{amstext}
\usepackage{amssymb}
\usepackage{graphicx}

\usepackage{xr}
\makeatletter
\newcommand*{\addFileDependency}[1]{
  \typeout{(#1)}
  \@addtofilelist{#1}
  \IfFileExists{#1}{}{\typeout{No file #1.}}
}
\makeatother

\newcommand*{\myexternaldocument}[1]{
    \externaldocument{#1}
    \addFileDependency{#1.tex}
    \addFileDependency{#1.aux}
}

\myexternaldocument{supplement}

\usepackage{chemformula}
\usepackage{braket}
\usepackage{float}
\usepackage{siunitx}
\usepackage{color}

\makeatletter

\ProvideTextCommand{\~}{LGR}[1]{\char126#1}

\@ifundefined{textcolor}{}
{%
 \definecolor{BLACK}{gray}{0}
 \definecolor{WHITE}{gray}{1}
 \definecolor{RED}{rgb}{1,0,0}
 \definecolor{GREEN}{rgb}{0,1,0}
 \definecolor{BLUE}{rgb}{0,0,1}
 \definecolor{CYAN}{cmyk}{1,0,0,0}
 \definecolor{MAGENTA}{cmyk}{0,1,0,0}
 \definecolor{YELLOW}{cmyk}{0,0,1,0}
}

\usepackage[english]{babel}
\def\urlprefix{}
\def\url#1{}
\addto\captionsenglish{%
}

\usepackage[none]{hyphenat}

\def\eV{\,\textrm{eV}}
\def\meV{\,\textrm{meV}}

\def\Mn{\,\textrm{Mn}}
\def\Ce{\,\textrm{Ce}}
\def\O{\,\textrm{O}}

\newcommand{\angstrom}{\,\mbox{\normalfont\AA}}

\usepackage{siunitx}

\usepackage{babel}
\usepackage{babel}
\usepackage{babel}

\makeatother

\usepackage{babel}

\begin{document}

\title{Long-Range Magnetic Exchange Pathways in Complex Clusters from First-Principles}
\author{Dian-Teng Chen}
\thanks{These two authors contributed equally}
\affiliation{Department of Physics and the Quantum Theory Project, University of Florida, Gainesville, Florida
32611, USA}

\author{Jia Chen}
\thanks{These two authors contributed equally}
\affiliation{Department of Physics and the Quantum Theory Project, University of Florida, Gainesville, Florida
32611, USA}

\author{Xiang-Guo Li}
\affiliation{Department of Physics and the Quantum Theory Project, University of Florida, Gainesville, Florida
32611, USA}


\author{George Christou}
\affiliation{Department of Chemistry, University of Florida, Gainesville, Florida
32611, USA}




\author{Xiao-Guang Zhang}
\affiliation{Department of Physics and the Quantum Theory Project, University of Florida, Gainesville, Florida
32611, USA}

\author{Hai-Ping Cheng}
 \email{hping@ufl.edu}
\affiliation{Department of Physics and the Quantum Theory Project, University of Florida, Gainesville, Florida
32611, USA}

\begin{abstract}

This work builds a bridge between density functional theory (DFT) and model interpretations of Anderson's superexchange theory by constructing a $f$-$d$-$p$  model with DFT Wannier functions to enable a direct quantum many-body solution within an embedding approach.
When applied to long-range magnetic interactions in a Mn-Ce magnetic molecule, we are able to obtain numerical insights about 
double exchange and superexchange interactions. Direct metal-metal charge transfer processes are generally weak in this molecule, which leads to small contributions from double exchange interactions. For long-range interactions, Mn-Ce charge transfer is not significant compared to Ce-O charge transfer. The unusual superexchange between Mn atoms with different valence states is identified as the dominant mechanism.  
This procedure opens a path for quantitative understanding of different exchange interactions in complex magnetic systems,  including molecular magnets, transition metal organic frameworks, and other solid materials. 

\end{abstract}

\maketitle


Because of national interest in quantum information science \cite{RN3276}, magnetic molecules (MM) are back in competition and in turn bring their unique challenges to the theory of magnetism.
Aside from a few single-ion examples, MMs usually contain multiple magnetic transition metal ions whose magnetic interactions can present rich and complex physics, 
especially when these ions are of mixed valence states such as the  $ \Mn_m \Ce_n$ oxo clusters and the well known SMM \ch{[Mn12O12(O2CR)16(H2O)4]}  \cite{caneschi1991alternating,sessoli1993high,sessoli1993magnetic}.  Molecular $ \Mn_m \Ce_n$ oxo clusters, first reported in 2003 \cite{RN3273}, have been synthesized  recently to study a variety of magnetic interactions involving $f$-electrons \cite{thuijs2017molecular,das2020long}. 
Analogies to bulk perovskite manganites are drawn  \cite{thuijs2017molecular}, where magnetic interactions are rich and complex and later long-range ferromagnetic couplings are observed  \cite{das2020long}.

For magnetic insulators, Anderson's superexchange theory \cite{Anderson1959} has been the guiding principle for the understanding of exchange couplings. In conductive compounds, the double exchange process \cite{Zener1951} is often the relevant coupling mechanism, particularly in mixed-valence compounds \cite{Coey1999}. For MMs, a complex molecular structures and the large number of orbitals involved often make identifying the exchange mechanism a difficult task. Applying semi-empirical rules like the Goodenough-Kanamori rule \cite{Goodenough1955, Goodenough1958, Kanamori1959} comes with complications when bond angles deviate from \ang{90} or \ang{180}. Formulae for exchange couplings $J$ derived from simplified models \cite{Fazekas1999} are not appropriate for systems with multiple on-site $d$ or $f$ orbitals and crystal fields with low symmetry. 
Complementary to simple models and empirical rules are first-principles based approaches, among which density functional theory  (DFT) \cite{kohn1965self} with improved functionals \cite{perdew1996generalized} and computational efficiency becomes the standard tool. Nevertheless, it is not always easy to quantitatively characterize exchange couplings directly using results from DFT calculations. 

\begin{figure}
\includegraphics[width=1.0\columnwidth]{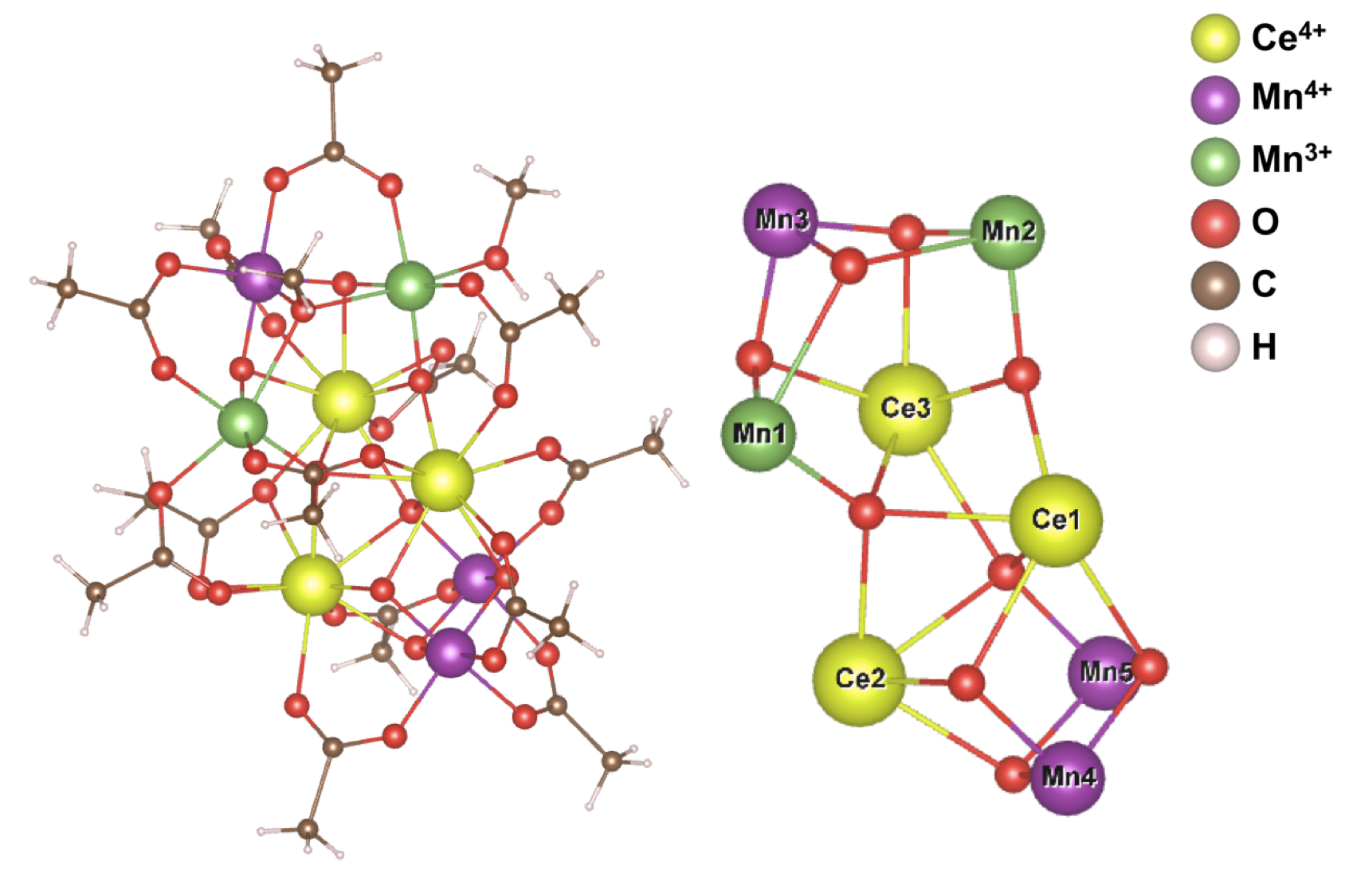}
\caption{Molecular structure of \ch{Mn5Ce3} with its core region displayed on the right.}
\label{fig:Mn5Ce3_1_structure}
\end{figure}

\begin{figure}
\includegraphics[width=0.8\columnwidth]{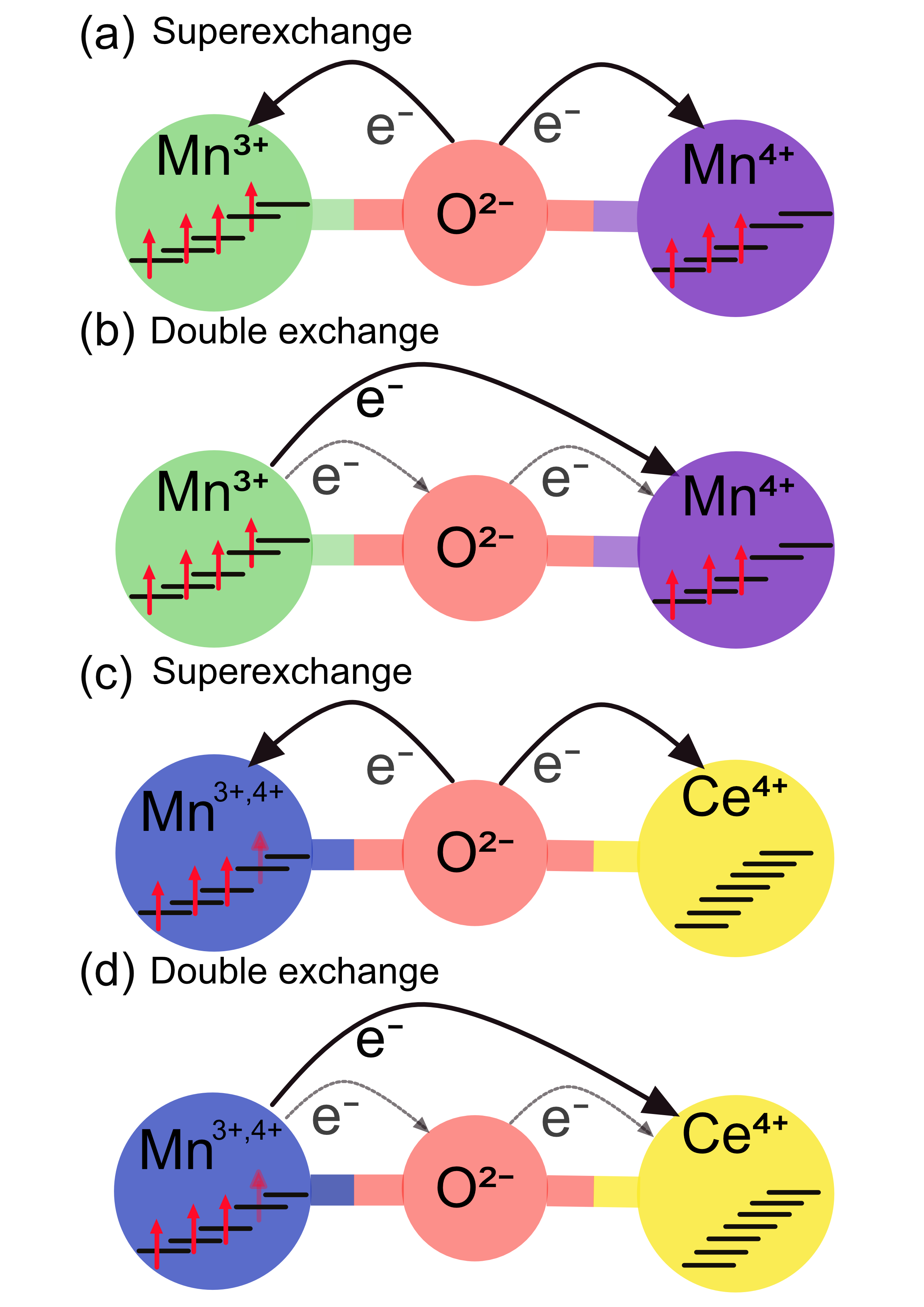}
\caption{
Panel (a): superexchange process between $\Mn^{3+}$ and $\Mn^{4+}$.
Panel (b): double exchange process between $\Mn^{3+}$ and $\Mn^{4+}$. 
Panel (c): superexchange process between $\Mn^{3+, 4+}$ and $\Ce^{4+}$. 
Panel (d): double exchange process between $\Mn^{3+, 4+}$ and $\Ce^{4+}$. }
\label{cartoon}
\end{figure}

In this work, we put forth a general approach to quantify various magnetic coupling strengths, e.g. double exchange versus superexchange. This approach combines DFT and many-body theory utilizing a quantum embedding scheme to quantitatively determine the mechanisms for exchange coupling. While our method is applicable to a wide range of magnetic clusters, our immediate interest lies in addressing issues emerging from studies of $ \Mn_5 \Ce_3$ clusters.  
The \ch{Mn5Ce3} complex (molecular structure shown in Fig.~\ref{fig:Mn5Ce3_1_structure}) is experimentally determined to be ferromagnetic, and this is supported by DFT+$U$ calculations \cite{DasGupta2020}.
Two important questions about the magnetic coupling remain unanswered. First, whether the two observed valance states of Mn (+3 and +4) have an impact on coupling. Second, what is the role of empty Ce $f$ orbitals in the long-range magnetic interaction and how can one quantify it. (two exchange processes are shown in Fig.~\ref{cartoon}. Insights obtained from this work, 
empowered by the joint DFT-quantum embedding method, nail down these unanswered questions and open up possibilities for a fresh new look at a wider range of old and new MMs. 

Quantum embedding methods, such as dynamical mean-field theory \cite{Georges1996, Kotliar2006}, density matrix embedding \cite{Knizia2012}, and self-energy embedding theory \cite{Lan2015}, have been very successful for materials with strong electron-electron interactions, and especially for molecules. SMMs with $d$ or $f$ electrons are a particularly good target for such a theoretical approach. 

Step I in our approach is to perform first-principles calculations of the molecule, including core and ligands. This step is divided into two sets of calculations: the first is to obtain electronic and magnetic properties of the Mn-Ce compounds and the second is to prepare the electron hopping matrix for many body treatment.
Both are done within the frame-work of Kohn-Sham density functional theory (DFT) \cite{kohn1965self} using the Perdew-Burke-Ernzerhof (PBE) exchange correlation functional \cite{perdew1996generalized} and  projector-augmented-wave (PAW) pseudopotentials \cite{blochl1994projector,kresse1999ultrasoft} in  conjunction with a plane-wave basis as implemented in the Vienna Ab-initio Simulation Package (VASP) \cite{kresse1996efficiency,kresse1996efficient}. Single Gamma point is used with a plane-wave cutoff energy of $500\eV$ and self-consistent convergence criterion $10^{-5}\eV$ . The first set of calculations is spin polarized with finite Hubbard $U$  added to Mn-$d$ and Ce-$f$ orbitals; and the second set is spin non-polarized with zero $U$.

Following step I, two parallel analyses are developed, each providing information necessary for understanding complex magnetic clusters. To study the exchange coupling strength, we optimize the structure of the MM and calculate energies for all possible magnetic configurations using a spin-polarized DFT+$U$ scheme. We then adopt a Heisenberg model
\begin{equation}
    H=-\sum_{i<j}J_{ij}  \, \vec{S_i} \cdot \vec{S_j}
\label{eq:H}
\end{equation}
where $\vec{S_i}$ and $\vec{S_j}$ are the spins of two interacting Manganese atoms and $J_{ij}$ is their exchange coupling constant. 
The broken symmetry approach formula \cite{ruiz1999broken} is used to extract the pairwise exchange parameter $J$, 
\begin{equation}
    E_\textrm{HS}-E_\textrm{BS}=-(2S_1S_2+S_2)/J_{12} ,
\label{eq:BS}
\end{equation}
where $S_1$ and $S_2$ are the two interacting local spins ($S_1 \geq S_2$) and $J_{12}$ is their exchange coupling constant. Here $E_\textrm{HS}$ is the high spin state energy of both spins in the same direction and $E_\textrm{BS}$ is the broken symmetry state energy of the two spins being in opposite directions. 

\begin{figure}
\includegraphics[width=0.5\columnwidth]{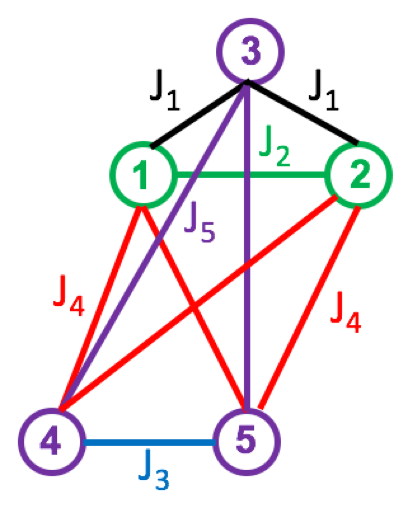}
\caption{The schematic diagram of the exchange interactions between the Mn atoms in \ch{Mn5Ce3}. Green circles represent \ch{Mn^3+} ions ($S=2$) and purple circles represent \ch{Mn^4+} ions ($S=3/2$).
Different line colors represent different $J$'s. Distances between Mn atoms (by the order of $J_1$--$J_5$) are: $d_{23} = 2.94 \angstrom $, $d_{13} = 3.02 \angstrom$, $d_{12} =  4.15 \angstrom $, $d_{45} = 2.74 \angstrom $, $d_{25} = 6.06 \angstrom $, $d_{14} = 5.89 \angstrom $, $d_{15} = 5.99 \angstrom $, $d_{24} = 7.00 \angstrom $, $d_{34} = 8.04 \angstrom $, $d_{35} = 7.22 \angstrom$ . The ground state is a FM state with $ S=17/2$.}
\label{fig:Mn5Ce3_1_5j}
\end{figure}

The calculated structure of the \ch{Mn5Ce3} molecule is shown in Fig.~\ref{fig:Mn5Ce3_1_structure}. The five Manganese ions are separated into two groups by the three Ce ions. The Mn ions within each group have strong interactions with each other, while the interactions between two Mn ions of different groups are much weaker but not negligible. Because of low symmetry there are, in principle, eight different $J$ (see supplement Fig.~\ref{fig:Mn5Ce3_1_8j}). 
As an approximation, we reduce the number to five distinct $J$'s (Fig.~\ref{fig:Mn5Ce3_1_5j}). $J_1$ is the interaction between \ch{Mn^3+} and \ch{Mn^4+}, $J_2$ is the interaction between two \ch{Mn^3+} and $J_3$ is the interaction between two \ch{Mn^4+} within the two groups respectively.
$J_4$ and $J_5$ are interactions between Mn ions from two different groups, with same or different valence. The fitting results of these constants (Table~\ref{table:Mn5Ce3_1_5j}) show
that all the interaction between two Mn ions are ferromagnetic (FM) except the interaction  $J_2$ between Mn1 and Mn2, which turns out the be anti-ferromagnetic (AFM). The AFM coupling $J_2$ is compensated by the much stronger FM interaction $J_1$ between Mn1 and Mn3, Mn2 and Mn3, and the ground state of \ch{Mn5Ce3} remains a FM state, as both observed by experiment and obtained from DFT+$U$ calculations 
(Supplement Table~\ref{table:Mn5Ce3_1_configs}). DFT calculations give same relative strength for $J$'s compared to experiments. The interaction between the two bottom \ch{Mn^4+}, $J_3$, is the strongest because of the short distance between these two ions. As the Mn ions with $J_1$ interactions have different valence, one wonders if double exchange dominates $J_1$, which is a key question of this work and will be discussed in great detail later. 

\begin{table}
\centering
\begin{tabular}{|p{8mm}|p{8mm}|p{8mm}|p{8mm}|p{8mm}|p{8mm}|} 
 \hline
 $J$'s & $J_1$ & $J_2$ &$J_3$ &$J_4$ &$J_5$  \\
\hline
   DFT &  43.8 &  $-7.5$ &  48.7 &  1.2 &  1.0\\
\hline
   Expt & 22.7 & $-4.0$ & 52.4 & 0.5 & 0.1 \\
\hline
\end{tabular}
\caption{The fitted exchange coupling constants J ($\textrm{cm}^{-1}$) of \ch{Mn5Ce3} from first-principles total energy calculations and from the experimental susceptibility curve, using five different $J$'s 
(Fig.~\ref{fig:Mn5Ce3_1_5j}). Positive values mean ferromagnetic coupling.}
\label{table:Mn5Ce3_1_5j}
\end{table}

From the spin non-polarized calculations, Wannier functions (WF) can be generated from unitary transformations of Bloch waves in a given energy window. We adopt Maximally Localized Wannier functions (MLWFs) \cite{marzari2012maximally,marzari1997maximally} as the local basis for first-principles based model calculations, because they can not only reproduce observables from DFT calculations, but also simplify the effective Hamiltonian for the model system due to their resemblance to orthogonal atomic orbitals.  We calculate the matrix elements of the effective Hamiltonian using the Wannier functions as basis,
\begin{equation}
    H_{mn}=\bra{W_m}\hat{H}\ket{W_n}
\label{eq:WF}
\end{equation}



Based on those MLWFs, a $f$-$d$-$p$ model with hopping and local electron-electron interaction of $d$ and $f$ orbitals for the core of \ch{Mn5Ce3} is constructed. The Hamiltonian can be decomposed into three parts, hopping, electron-electron interaction, and double-counting, 
\begin{align}\label{tot_Ham}
\hat{H} = \hat{H}_{\textrm{hop}}+ \sum_{a=\Mn,\Ce} \hat{H}^{a}_{\textrm{int}} + \sum_{a=\Mn,\Ce} \hat{H}^{a}_{\textrm{dc}},
\end{align}
To account for the possibility of long range processes, the hopping Hamiltonian is not limited to near-neighbor terms. The diagonal terms of the hopping Hamiltonian are the  chemical potential of orbitals:
\begin{align}\label{hop_Ham}
\hat{H}_{\textrm{hop}} = \sum_{i,j,\sigma} t_{ij} \, c^{\dagger}_{i,\sigma} \, c_{j,\sigma},
\end{align}
where $i,j$ label MLWFs and $\sigma$
can be either $\uparrow$ or $\downarrow$ spins. Parameters $t_{ij}$ were obtained from the DFT Hamiltonian with MLWFs as basis, as in Eq.~\ref{eq:WF}

For $d$ or $f$ orbitals belong to one Mn or Ce ion, the on-site electron-electron interaction is modeled by the density-density part of Slater-Kanamori Hamiltonian \cite{Kanamori1963}
parameterized by intra-orbital Coulomb interaction $U$ and exchange interaction $J$:
\begin{eqnarray}
  \hat{H}_\textrm{int}^{a} &=& U\sum_{i}  \hat{n}_{ai\uparrow} \, \hat{n}_{ai\downarrow} 
  + (U-2J) \sum_{ i\neq j} \hat{n}_{ai\uparrow} \, \hat{n}_{aj\downarrow} \nonumber \\
  &&+ (U-3J)\sum_{i > j,\; \sigma}\hat{n}_{ai\sigma} \, \hat{n}_{aj\sigma} \label{int}
\end{eqnarray}
where $a$ labels Mn and Ce ions. For the double counting, which is used with the intention to cancel existing on-site interaction in DFT calculation, we use the spin-independent form given by Anisimov \textit{et al.} \cite{Anisimov1993},
\begin{align}\label{dcv}
 V_{\textrm{dc}}^{a} 
 = U \Bigl( N_{a} - \frac{1}{2} \Bigr) 
  - J \Bigl( \frac{N_{a}}{2} - \frac{1}{2} \Bigr), 
\end{align}
where $N_{a}$ is occupation number for one Mn or Ce ion, and the double-counting Hamiltonian is:
\begin{align}
    \hat{H}_{dc}^{a} = -V_{dc}^{a}\sum_{i} \hat{n}_{ai}
\end{align}

Superexchange and double exchange processes are both kinetic exchange in nature because electron hoppings are necessary in both cases. The key difference is that the superexchange process needs hopping between magnetic transition metal ions ($\Mn^{3+}$ and $\Mn^{4+}$ in this work) and diamagnetic bridging ions ($\O^{2-}$ here); however, the double exchange process demands electron transfer among magnetic ions, (Mn to Mn) which can be the result of direct hopping between magnetic ions, or higher order processes consisting of two or more hopping terms and bridging ions. 

For finite systems, all the exchange processes can be described by the configuration interaction (CI) method \cite{DAVIDSHERRILL1999}. Since we are mainly interested in the exchange mechanism, we choose excitations that are important to the exchange processes, making it a selected CI method. Total energies of the model system and relevance of each excitation can be obtained from diagonalization of the Hamiltonian. Insights about the mechanism for exchange couplings can be obtained by analyzing the CI wave-function, specifically comparing weights of excitations corresponding to different exchange processes.



\begin{figure}
\includegraphics[width=1.0\columnwidth]{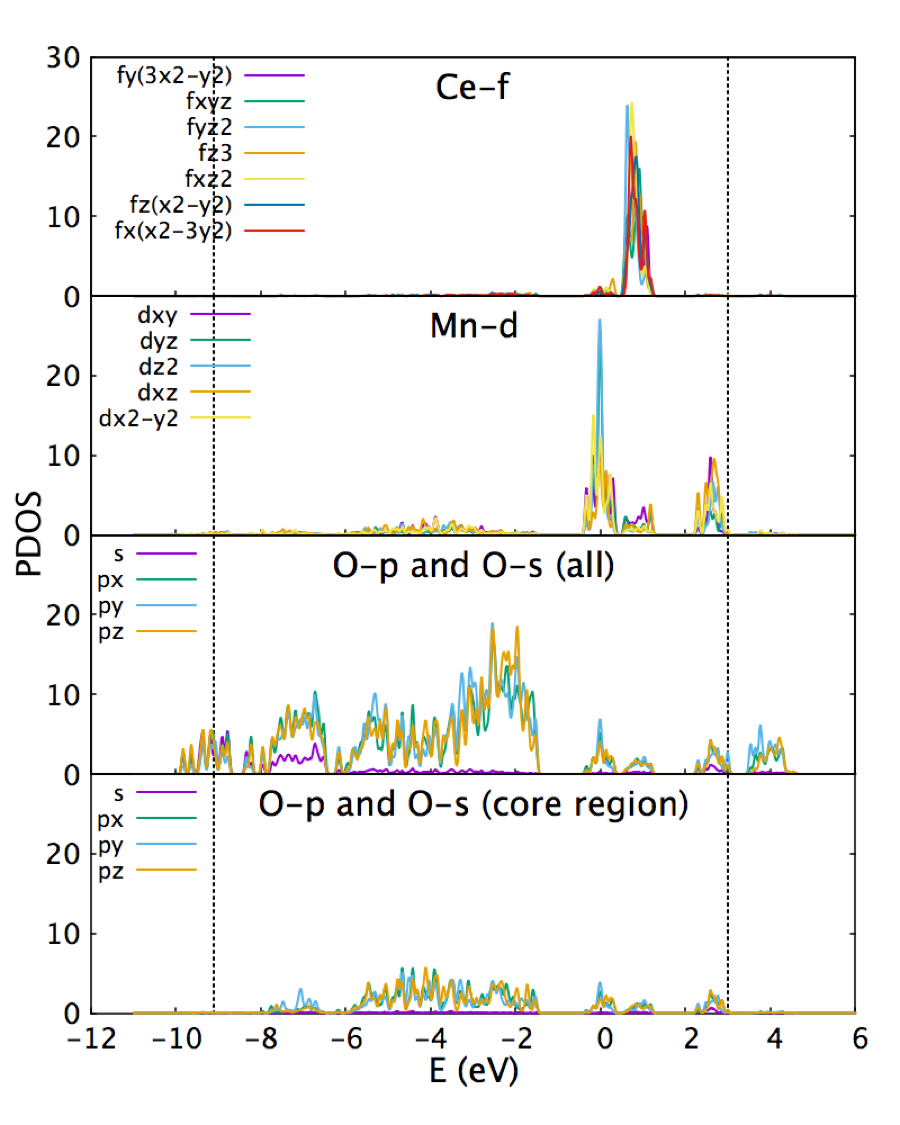}
\caption{Projected density of states (PDOS) of the Ce-$f$, Mn-$d$, O-$p$ and O-$s$ orbitals of \ch{Mn5Ce3}. The dotted lines bracket the energy window for constructing WFs. Fermi energy is set to be zero.}
\label{fig:Mn5Ce3_1_dos}
\end{figure}

To construct MLWFs for the \ch{Mn5Ce3} complex we use the Wannier90 code \cite{Mostofi2014}. We first examine the energy range of all the Ce-$f$, Mn-$d$, O-$p$ and O-$s$ orbitals in the core region of the molecule by plotting the PDOS of these orbitals (Fig.~\ref{fig:Mn5Ce3_1_dos}). The disentanglement energy is chosen to be large enough to include all these orbitals. For our molecule the energy range is from $-9.1$ to $3.0 \eV$ (Fermi energy is zero). A total of 190 orbitals (including 21 Ce-$f$, 25 Mn-$d$, 108 O-$p$ and 36 O-$s$ orbitals) are chosen as the projection seeds for the initial guess of the Wannier functions. In order to include all orbitals in the core region, some orbitals from neighboring O atoms are also included.

Based on MLWFs generated from restricted DFT calculations, we built a model for the core region 
$[\Mn^{\,3+}_2 \Mn^{\,4+}_3 \Ce^{\,4+}_3 \O^{\,2-}_9]$ 
of this molecule. All five $d$ orbitals for each manganese, seven $f$ orbitals for each cerium and four $s$ and $p$ orbitals for each oxygen, for a total of 164 spin-orbitals, are included in the model, with 89 of them occupied. We incorporated all the filled oxygen orbitals and empty cerium orbitals for the model to encompass all possible exchange pathways of interest.

The reference state in the CI calculation (Suppl.~IV) was constructed as the ferromagnetic symmetry breaking state with charge and spin state for each ion as identified experimentally \cite{DasGupta2020}. Excitations, which are relevant to exchange processes, were selected for the CI calculation. As discussed earlier, two charge states: $\Mn^{3+}$ and $\Mn^{4+}$ are in adjacency in this molecule. Double exchange processes were investigated by involving excitations that transfer one electron from $\Mn^{3+}$ to $\Mn^{4+}$, and 48 such excitations can be found. For the sake of completeness, 18 excitations that transfer one electron from $\Mn^{4+}$ to $\Mn^{3+}$ are also included in the calculation. One thing to notice is that those excitations can connect not only to the reference state, but also to other excitations.  To accommodate exchange processes, excitations that transfer electrons from $\O^{2-}$ to $\Mn^{3+,\,4+}$, for both majority and minority spins are included. To address the origin of long-range exchange interactions in this molecule, which couples Mn ions separated by Ce ions, excitations transfer one electron from $\O^{2-}$ to $\Ce^{4+}$ and from $\Mn^{3+,4+}$ to $\Ce^{4+}$ are also included. One question we would like to answer is whether the long-range exchange interaction comes from a direct metal-metal interaction, (Mn-Ce for our case) or a connected superexchange pathway via bridging $\O^{2-}$.

\begin{center}
\begin{figure}[h]
\includegraphics[width=\columnwidth]{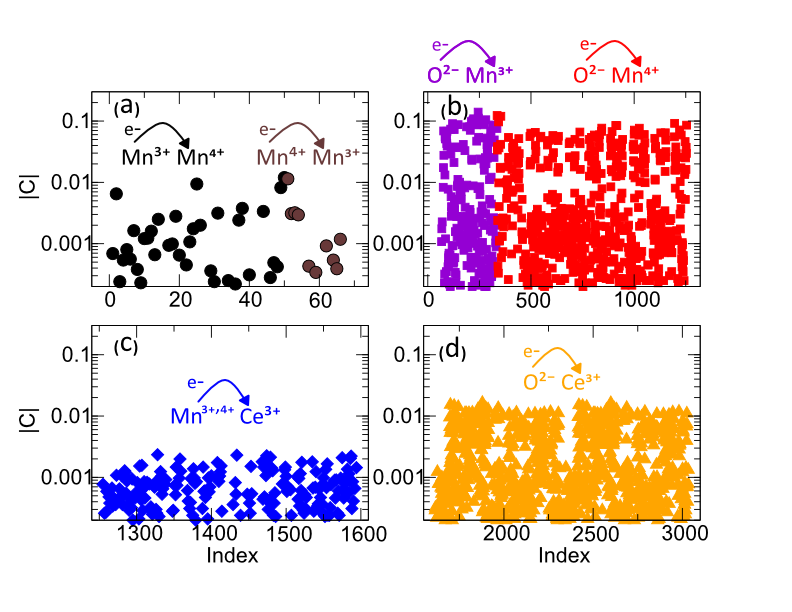}
\caption{Absolute value of coefficients for excitations obtained from diagonalization, plotted on a logarithmic scale. 
Panel (a): Coefficients for excitations that transfer electrons between $\Mn^{3+}$ and $\Mn^{4+}$.
Panel (b): Coefficients for excitations that transfer electrons from $\O^{2-}$ to $\Mn^{3+,4+}$. 
Panel (c): Coefficients for excitations that transfer electrons from $\Mn^{3+,\,4+}$ to $\Ce^{4+}$.
Panel (d): Coefficients for excitations that transfer electrons from $\O^{2-}$ to $\Ce^{4+}$}
\label{coeff}
\end{figure}
\end{center}

Relative importances of these excitations were obtained by diagonalization of the Hamiltonian in Eq.~\ref{tot_Ham}. The coefficient for the reference state is 0.68 and occupation numbers for $\Mn^{3+}$, $\Mn^{4+}$ and $\Ce^{4+}$ are 4.08 and 3.12, and 0.006 respectively. The CI wave-function is consistent with experimentally observed charge and spin states. Absolute values of coefficients for excitations are found in Fig.~\ref{coeff}. We can see that the most important excitations for this molecule correspond to charge transfer from $\O^{2-}$ to $\Mn^{3+,\,4+}$. Not only they are large in number, their coefficients are also much larger (by about one order of magnitude) than those corresponding to charge transfer between Mn ions. Because no excitatons that transfer electrons from $\Mn^{3+}$ to $\Mn{4+}$ have significant coefficients in the CI wave-function, we conclude that double exchange is not important for exchange couplings in this molecule. Concerning long-range ferromagnetic interactions, from the small occupation number of $\Ce^{4+}$ (0.006), we can infer that $f$ orbitals do have a role in exchange interaction, but mostly just as empty orbitals. Secondly, the metal-metal interactions (Mn-Ce charge transfer) are week, since excitations of those processes have small coefficients. The exchange pathway for the long-range magnetic interaction is paved by diamagnetic $\O^{2-}$ ions, since both Mn-O and Ce-O charge transfer have much larger weights in the CI wave-function. 

We further estimate the contribution of double exchange by the changes of ground state eigenvalues when the metal-metal charge transfer excitations are excluded.  When Mn1, Mn2  to Mn3 (see Fig.~\ref{fig:Mn5Ce3_1_structure} for structure) charge transfers are excluded, the total energy increases by $16.0 \, \textrm{cm}^{-1}$. When taking the experimental value of $J_1 = 22.7 \,  \textrm{cm}^{-1}$, energy decrease due to near neighboring Mn$^{3+}$-Mn$^{4+}$ charge transfers account for  about 12\% of the experimental total exchange coupling. For the long-range coupling, the experimental values for $J_4$ and $J_5$ give an energy difference between spin configurations smaller than $1\meV$. 


\onecolumngrid


\twocolumngrid

To conclude, we have devised a quantum embedding scheme built on DFT and a selected CI method to quantify the relative importance of different exchange mechanisms. In particular we obtain from a diagonalization procedure coefficients for excitations corresponding to various charge transfer processes, turning previous phenomenological models of exchange interactions into a predictive theory. Our analysis shows that double exchange processes are not significant in \ch{Mn5Ce3}, and 
direct metal-metal interactions also are not crucial to the observed long-range interactions. Magnetic interactions in this molecule arise mainly from super-exchange via bridging oxygen, despite the mixed valence. The long-range FM interaction thus emerges from a collection of local superexchange processes. This quantum embedding scheme is also applied to the SMM $\Mn_3$ (Suppl.~Fig.~S2). A variety of MMs molecules will be examined in a future paper.

\section{Acknowledgment}

This work is supported as part of the Center for Molecular Magnetic Quantum Materials, an Energy Frontier Research Center funded by the U.S. Department of Energy, Office of Science, Basic Energy Sciences under Award No. DE-SC0019330. 
Computations were performed at NERSC
and University of Florida Research Computer Center.

\bibliography{References}

\begin{thebibliography}{32}%
\makeatletter
\providecommand \@ifxundefined [1]{%
 \@ifx{#1\undefined}
}%
\providecommand \@ifnum [1]{%
 \ifnum #1\expandafter \@firstoftwo
 \else \expandafter \@secondoftwo
 \fi
}%
\providecommand \@ifx [1]{%
 \ifx #1\expandafter \@firstoftwo
 \else \expandafter \@secondoftwo
 \fi
}%
\providecommand \natexlab [1]{#1}%
\providecommand \enquote  [1]{``#1''}%
\providecommand \bibnamefont  [1]{#1}%
\providecommand \bibfnamefont [1]{#1}%
\providecommand \citenamefont [1]{#1}%
\providecommand \href@noop [0]{\@secondoftwo}%
\providecommand \href [0]{\begingroup \@sanitize@url \@href}%
\providecommand \@href[1]{\@@startlink{#1}\@@href}%
\providecommand \@@href[1]{\endgroup#1\@@endlink}%
\providecommand \@sanitize@url [0]{\catcode `\\12\catcode `\$12\catcode
  `\&12\catcode `\#12\catcode `\^12\catcode `\_12\catcode `\%12\relax}%
\providecommand \@@startlink[1]{}%
\providecommand \@@endlink[0]{}%
\providecommand \url  [0]{\begingroup\@sanitize@url \@url }%
\providecommand \@url [1]{\endgroup\@href {#1}{\urlprefix }}%
\providecommand \urlprefix  [0]{URL }%
\providecommand \Eprint [0]{\href }%
\providecommand \doibase [0]{http://dx.doi.org/}%
\providecommand \selectlanguage [0]{\@gobble}%
\providecommand \bibinfo  [0]{\@secondoftwo}%
\providecommand \bibfield  [0]{\@secondoftwo}%
\providecommand \translation [1]{[#1]}%
\providecommand \BibitemOpen [0]{}%
\providecommand \bibitemStop [0]{}%
\providecommand \bibitemNoStop [0]{.\EOS\space}%
\providecommand \EOS [0]{\spacefactor3000\relax}%
\providecommand \BibitemShut  [1]{\csname bibitem#1\endcsname}%
\let\auto@bib@innerbib\@empty
\bibitem [{\citenamefont {Nielsen}\ and\ \citenamefont
  {Chuang}(2010)}]{RN3276}%
  \BibitemOpen
  \bibfield  {author} {\bibinfo {author} {\bibfnamefont {M.~A.}\ \bibnamefont
  {Nielsen}}\ and\ \bibinfo {author} {\bibfnamefont {I.~L.}\ \bibnamefont
  {Chuang}},\ }\href@noop {} {\emph {\bibinfo {title} {Quantum computation and
  quantum information}}},\ \bibinfo {edition} {new}\ ed.\ (\bibinfo
  {publisher} {Cambridge University Press},\ \bibinfo {address} {Cambridge ;
  New York},\ \bibinfo {year} {2010})\ pp.\ \bibinfo {pages} {xxxi, 676
  p.}\BibitemShut {Stop}%
\bibitem [{\citenamefont {Caneschi}\ \emph {et~al.}(1991)\citenamefont
  {Caneschi}, \citenamefont {Gatteschi}, \citenamefont {Sessoli}, \citenamefont
  {Barra}, \citenamefont {Brunel},\ and\ \citenamefont
  {Guillot}}]{caneschi1991alternating}%
  \BibitemOpen
  \bibfield  {author} {\bibinfo {author} {\bibfnamefont {A.}~\bibnamefont
  {Caneschi}}, \bibinfo {author} {\bibfnamefont {D.}~\bibnamefont {Gatteschi}},
  \bibinfo {author} {\bibfnamefont {R.}~\bibnamefont {Sessoli}}, \bibinfo
  {author} {\bibfnamefont {A.~L.}\ \bibnamefont {Barra}}, \bibinfo {author}
  {\bibfnamefont {L.~C.}\ \bibnamefont {Brunel}}, \ and\ \bibinfo {author}
  {\bibfnamefont {M.}~\bibnamefont {Guillot}},\ }\href@noop {} {\bibfield
  {journal} {\bibinfo  {journal} {Journal of the American Chemical Society}\
  }\textbf {\bibinfo {volume} {113}},\ \bibinfo {pages} {5873} (\bibinfo {year}
  {1991})}\BibitemShut {NoStop}%
\bibitem [{\citenamefont {Sessoli}\ \emph
  {et~al.}(1993{\natexlab{a}})\citenamefont {Sessoli}, \citenamefont {Tsai},
  \citenamefont {Schake}, \citenamefont {Wang}, \citenamefont {Vincent},
  \citenamefont {Folting}, \citenamefont {Gatteschi}, \citenamefont
  {Christou},\ and\ \citenamefont {Hendrickson}}]{sessoli1993high}%
  \BibitemOpen
  \bibfield  {author} {\bibinfo {author} {\bibfnamefont {R.}~\bibnamefont
  {Sessoli}}, \bibinfo {author} {\bibfnamefont {H.~L.}\ \bibnamefont {Tsai}},
  \bibinfo {author} {\bibfnamefont {A.~R.}\ \bibnamefont {Schake}}, \bibinfo
  {author} {\bibfnamefont {S.}~\bibnamefont {Wang}}, \bibinfo {author}
  {\bibfnamefont {J.~B.}\ \bibnamefont {Vincent}}, \bibinfo {author}
  {\bibfnamefont {K.}~\bibnamefont {Folting}}, \bibinfo {author} {\bibfnamefont
  {D.}~\bibnamefont {Gatteschi}}, \bibinfo {author} {\bibfnamefont
  {G.}~\bibnamefont {Christou}}, \ and\ \bibinfo {author} {\bibfnamefont
  {D.~N.}\ \bibnamefont {Hendrickson}},\ }\href@noop {} {\bibfield  {journal}
  {\bibinfo  {journal} {Journal of the American Chemical Society}\ }\textbf
  {\bibinfo {volume} {115}},\ \bibinfo {pages} {1804} (\bibinfo {year}
  {1993}{\natexlab{a}})}\BibitemShut {NoStop}%
\bibitem [{\citenamefont {Sessoli}\ \emph
  {et~al.}(1993{\natexlab{b}})\citenamefont {Sessoli}, \citenamefont
  {Gatteschi}, \citenamefont {Caneschi},\ and\ \citenamefont
  {Novak}}]{sessoli1993magnetic}%
  \BibitemOpen
  \bibfield  {author} {\bibinfo {author} {\bibfnamefont {R.}~\bibnamefont
  {Sessoli}}, \bibinfo {author} {\bibfnamefont {D.}~\bibnamefont {Gatteschi}},
  \bibinfo {author} {\bibfnamefont {A.}~\bibnamefont {Caneschi}}, \ and\
  \bibinfo {author} {\bibfnamefont {M.}~\bibnamefont {Novak}},\ }\href@noop {}
  {\bibfield  {journal} {\bibinfo  {journal} {Nature}\ }\textbf {\bibinfo
  {volume} {365}},\ \bibinfo {pages} {141} (\bibinfo {year}
  {1993}{\natexlab{b}})}\BibitemShut {NoStop}%
\bibitem [{\citenamefont {Tasiopoulos}\ \emph {et~al.}(2003)\citenamefont
  {Tasiopoulos}, \citenamefont {Wernsdorfer}, \citenamefont {Moulton},
  \citenamefont {Zaworotko},\ and\ \citenamefont {Christou}}]{RN3273}%
  \BibitemOpen
  \bibfield  {author} {\bibinfo {author} {\bibfnamefont {A.~J.}\ \bibnamefont
  {Tasiopoulos}}, \bibinfo {author} {\bibfnamefont {W.}~\bibnamefont
  {Wernsdorfer}}, \bibinfo {author} {\bibfnamefont {B.}~\bibnamefont
  {Moulton}}, \bibinfo {author} {\bibfnamefont {M.~J.}\ \bibnamefont
  {Zaworotko}}, \ and\ \bibinfo {author} {\bibfnamefont {G.}~\bibnamefont
  {Christou}},\ }\href {\doibase DOI 10.1021/ja0385134} {\bibfield  {journal}
  {\bibinfo  {journal} {Journal of the American Chemical Society}\ }\textbf
  {\bibinfo {volume} {125}},\ \bibinfo {pages} {15274} (\bibinfo {year}
  {2003})}\BibitemShut {NoStop}%
\bibitem [{\citenamefont {Thuijs}\ \emph {et~al.}(2017)\citenamefont {Thuijs},
  \citenamefont {Li}, \citenamefont {Wang}, \citenamefont {Abboud},
  \citenamefont {Zhang}, \citenamefont {Cheng},\ and\ \citenamefont
  {Christou}}]{thuijs2017molecular}%
  \BibitemOpen
  \bibfield  {author} {\bibinfo {author} {\bibfnamefont {A.~E.}\ \bibnamefont
  {Thuijs}}, \bibinfo {author} {\bibfnamefont {X.-G.}\ \bibnamefont {Li}},
  \bibinfo {author} {\bibfnamefont {Y.-P.}\ \bibnamefont {Wang}}, \bibinfo
  {author} {\bibfnamefont {K.~A.}\ \bibnamefont {Abboud}}, \bibinfo {author}
  {\bibfnamefont {X.-G.}\ \bibnamefont {Zhang}}, \bibinfo {author}
  {\bibfnamefont {H.-P.}\ \bibnamefont {Cheng}}, \ and\ \bibinfo {author}
  {\bibfnamefont {G.}~\bibnamefont {Christou}},\ }\href@noop {} {\bibfield
  {journal} {\bibinfo  {journal} {Nature communications}\ }\textbf {\bibinfo
  {volume} {8}},\ \bibinfo {pages} {500} (\bibinfo {year} {2017})}\BibitemShut
  {NoStop}%
\bibitem [{\citenamefont {Das~Gupta}\ \emph
  {et~al.}(2020{\natexlab{a}})\citenamefont {Das~Gupta}, \citenamefont
  {Stewart}, \citenamefont {Chen}, \citenamefont {Abboud}, \citenamefont
  {Cheng}, \citenamefont {Hill},\ and\ \citenamefont {Christou}}]{das2020long}%
  \BibitemOpen
  \bibfield  {author} {\bibinfo {author} {\bibfnamefont {S.}~\bibnamefont
  {Das~Gupta}}, \bibinfo {author} {\bibfnamefont {R.~L.}\ \bibnamefont
  {Stewart}}, \bibinfo {author} {\bibfnamefont {D.-T.}\ \bibnamefont {Chen}},
  \bibinfo {author} {\bibfnamefont {K.~A.}\ \bibnamefont {Abboud}}, \bibinfo
  {author} {\bibfnamefont {H.-P.}\ \bibnamefont {Cheng}}, \bibinfo {author}
  {\bibfnamefont {S.}~\bibnamefont {Hill}}, \ and\ \bibinfo {author}
  {\bibfnamefont {G.}~\bibnamefont {Christou}},\ }\href@noop {} {\bibfield
  {journal} {\bibinfo  {journal} {Inorganic Chemistry}\ }\textbf {\bibinfo
  {volume} {59}},\ \bibinfo {pages} {8716} (\bibinfo {year}
  {2020}{\natexlab{a}})}\BibitemShut {NoStop}%
\bibitem [{\citenamefont {Anderson}(1959)}]{Anderson1959}%
  \BibitemOpen
  \bibfield  {author} {\bibinfo {author} {\bibfnamefont {P.~W.}\ \bibnamefont
  {Anderson}},\ }\href {\doibase 10.1103/PhysRev.115.2} {\bibfield  {journal}
  {\bibinfo  {journal} {Phys. Rev.}\ }\textbf {\bibinfo {volume} {115}},\
  \bibinfo {pages} {2} (\bibinfo {year} {1959})}\BibitemShut {NoStop}%
\bibitem [{\citenamefont {Zener}(1951)}]{Zener1951}%
  \BibitemOpen
  \bibfield  {author} {\bibinfo {author} {\bibfnamefont {C.}~\bibnamefont
  {Zener}},\ }\href {\doibase 10.1103/PhysRev.82.403} {\bibfield  {journal}
  {\bibinfo  {journal} {Phys. Rev.}\ }\textbf {\bibinfo {volume} {82}},\
  \bibinfo {pages} {403} (\bibinfo {year} {1951})}\BibitemShut {NoStop}%
\bibitem [{\citenamefont {Coey}\ \emph {et~al.}(1999)\citenamefont {Coey},
  \citenamefont {Viret},\ and\ \citenamefont {von Molnár}}]{Coey1999}%
  \BibitemOpen
  \bibfield  {author} {\bibinfo {author} {\bibfnamefont {J.~M.~D.}\
  \bibnamefont {Coey}}, \bibinfo {author} {\bibfnamefont {M.}~\bibnamefont
  {Viret}}, \ and\ \bibinfo {author} {\bibfnamefont {S.}~\bibnamefont {von
  Molnár}},\ }\href {\doibase 10.1080/000187399243455} {\bibfield  {journal}
  {\bibinfo  {journal} {Advances in Physics}\ }\textbf {\bibinfo {volume}
  {48}},\ \bibinfo {pages} {167} (\bibinfo {year} {1999})}\BibitemShut
  {NoStop}%
\bibitem [{\citenamefont {Goodenough}(1955)}]{Goodenough1955}%
  \BibitemOpen
  \bibfield  {author} {\bibinfo {author} {\bibfnamefont {J.~B.}\ \bibnamefont
  {Goodenough}},\ }\href {\doibase 10.1103/PhysRev.100.564} {\bibfield
  {journal} {\bibinfo  {journal} {Phys. Rev.}\ }\textbf {\bibinfo {volume}
  {100}},\ \bibinfo {pages} {564} (\bibinfo {year} {1955})}\BibitemShut
  {NoStop}%
\bibitem [{\citenamefont {Goodenough}(1958)}]{Goodenough1958}%
  \BibitemOpen
  \bibfield  {author} {\bibinfo {author} {\bibfnamefont {J.~B.}\ \bibnamefont
  {Goodenough}},\ }\href {\doibase
  https://doi.org/10.1016/0022-3697(58)90107-0} {\bibfield  {journal} {\bibinfo
   {journal} {Journal of Physics and Chemistry of Solids}\ }\textbf {\bibinfo
  {volume} {6}},\ \bibinfo {pages} {287 } (\bibinfo {year} {1958})}\BibitemShut
  {NoStop}%
\bibitem [{\citenamefont {Kanamori}(1959)}]{Kanamori1959}%
  \BibitemOpen
  \bibfield  {author} {\bibinfo {author} {\bibfnamefont {J.}~\bibnamefont
  {Kanamori}},\ }\href {\doibase https://doi.org/10.1016/0022-3697(59)90061-7}
  {\bibfield  {journal} {\bibinfo  {journal} {Journal of Physics and Chemistry
  of Solids}\ }\textbf {\bibinfo {volume} {10}},\ \bibinfo {pages} {87 }
  (\bibinfo {year} {1959})}\BibitemShut {NoStop}%
\bibitem [{\citenamefont {Fazekas}(1999)}]{Fazekas1999}%
  \BibitemOpen
  \bibfield  {author} {\bibinfo {author} {\bibfnamefont {P.}~\bibnamefont
  {Fazekas}},\ }\href {\doibase 10.1142/2945} {\emph {\bibinfo {title} {Lecture
  Notes on Electron Correlation and Magnetism}}}\ (\bibinfo  {publisher} {WORLD
  SCIENTIFIC},\ \bibinfo {year} {1999})\ \Eprint
  {http://arxiv.org/abs/https://www.worldscientific.com/doi/pdf/10.1142/2945}
  {https://www.worldscientific.com/doi/pdf/10.1142/2945} \BibitemShut {NoStop}%
\bibitem [{\citenamefont {Kohn}\ and\ \citenamefont
  {Sham}(1965)}]{kohn1965self}%
  \BibitemOpen
  \bibfield  {author} {\bibinfo {author} {\bibfnamefont {W.}~\bibnamefont
  {Kohn}}\ and\ \bibinfo {author} {\bibfnamefont {L.~J.}\ \bibnamefont
  {Sham}},\ }\href@noop {} {\bibfield  {journal} {\bibinfo  {journal} {Physical
  review}\ }\textbf {\bibinfo {volume} {140}},\ \bibinfo {pages} {A1133}
  (\bibinfo {year} {1965})}\BibitemShut {NoStop}%
\bibitem [{\citenamefont {Perdew}\ \emph {et~al.}(1996)\citenamefont {Perdew},
  \citenamefont {Burke},\ and\ \citenamefont
  {Ernzerhof}}]{perdew1996generalized}%
  \BibitemOpen
  \bibfield  {author} {\bibinfo {author} {\bibfnamefont {J.~P.}\ \bibnamefont
  {Perdew}}, \bibinfo {author} {\bibfnamefont {K.}~\bibnamefont {Burke}}, \
  and\ \bibinfo {author} {\bibfnamefont {M.}~\bibnamefont {Ernzerhof}},\
  }\href@noop {} {\bibfield  {journal} {\bibinfo  {journal} {Physical review
  letters}\ }\textbf {\bibinfo {volume} {77}},\ \bibinfo {pages} {3865}
  (\bibinfo {year} {1996})}\BibitemShut {NoStop}%
\bibitem [{\citenamefont {Das~Gupta}\ \emph
  {et~al.}(2020{\natexlab{b}})\citenamefont {Das~Gupta}, \citenamefont
  {Stewart}, \citenamefont {Chen}, \citenamefont {Abboud}, \citenamefont
  {Cheng}, \citenamefont {Hill},\ and\ \citenamefont
  {Christou}}]{DasGupta2020}%
  \BibitemOpen
  \bibfield  {author} {\bibinfo {author} {\bibfnamefont {S.}~\bibnamefont
  {Das~Gupta}}, \bibinfo {author} {\bibfnamefont {R.~L.}\ \bibnamefont
  {Stewart}}, \bibinfo {author} {\bibfnamefont {D.-T.}\ \bibnamefont {Chen}},
  \bibinfo {author} {\bibfnamefont {K.~A.}\ \bibnamefont {Abboud}}, \bibinfo
  {author} {\bibfnamefont {H.-P.}\ \bibnamefont {Cheng}}, \bibinfo {author}
  {\bibfnamefont {S.}~\bibnamefont {Hill}}, \ and\ \bibinfo {author}
  {\bibfnamefont {G.}~\bibnamefont {Christou}},\ }\href {\doibase
  10.1021/acs.inorgchem.0c00332} {\bibfield  {journal} {\bibinfo  {journal}
  {Inorganic Chemistry}\ }\textbf {\bibinfo {volume} {59}},\ \bibinfo {pages}
  {8716} (\bibinfo {year} {2020}{\natexlab{b}})}\BibitemShut {NoStop}%
\bibitem [{\citenamefont {Georges}\ \emph {et~al.}(1996)\citenamefont
  {Georges}, \citenamefont {Kotliar}, \citenamefont {Krauth},\ and\
  \citenamefont {Rozenberg}}]{Georges1996}%
  \BibitemOpen
  \bibfield  {author} {\bibinfo {author} {\bibfnamefont {A.}~\bibnamefont
  {Georges}}, \bibinfo {author} {\bibfnamefont {G.}~\bibnamefont {Kotliar}},
  \bibinfo {author} {\bibfnamefont {W.}~\bibnamefont {Krauth}}, \ and\ \bibinfo
  {author} {\bibfnamefont {M.~J.}\ \bibnamefont {Rozenberg}},\ }\href {\doibase
  10.1103/RevModPhys.68.13} {\bibfield  {journal} {\bibinfo  {journal} {Rev.
  Mod. Phys.}\ }\textbf {\bibinfo {volume} {68}},\ \bibinfo {pages} {13}
  (\bibinfo {year} {1996})}\BibitemShut {NoStop}%
\bibitem [{\citenamefont {Kotliar}\ \emph {et~al.}(2006)\citenamefont
  {Kotliar}, \citenamefont {Savrasov}, \citenamefont {Haule}, \citenamefont
  {Oudovenko}, \citenamefont {Parcollet},\ and\ \citenamefont
  {Marianetti}}]{Kotliar2006}%
  \BibitemOpen
  \bibfield  {author} {\bibinfo {author} {\bibfnamefont {G.}~\bibnamefont
  {Kotliar}}, \bibinfo {author} {\bibfnamefont {S.~Y.}\ \bibnamefont
  {Savrasov}}, \bibinfo {author} {\bibfnamefont {K.}~\bibnamefont {Haule}},
  \bibinfo {author} {\bibfnamefont {V.~S.}\ \bibnamefont {Oudovenko}}, \bibinfo
  {author} {\bibfnamefont {O.}~\bibnamefont {Parcollet}}, \ and\ \bibinfo
  {author} {\bibfnamefont {C.~A.}\ \bibnamefont {Marianetti}},\ }\href
  {\doibase 10.1103/RevModPhys.78.865} {\bibfield  {journal} {\bibinfo
  {journal} {Rev. Mod. Phys.}\ }\textbf {\bibinfo {volume} {78}},\ \bibinfo
  {pages} {865} (\bibinfo {year} {2006})}\BibitemShut {NoStop}%
\bibitem [{\citenamefont {Knizia}\ and\ \citenamefont
  {Chan}(2012)}]{Knizia2012}%
  \BibitemOpen
  \bibfield  {author} {\bibinfo {author} {\bibfnamefont {G.}~\bibnamefont
  {Knizia}}\ and\ \bibinfo {author} {\bibfnamefont {G.~K.-L.}\ \bibnamefont
  {Chan}},\ }\href {\doibase 10.1103/PhysRevLett.109.186404} {\bibfield
  {journal} {\bibinfo  {journal} {Phys. Rev. Lett.}\ }\textbf {\bibinfo
  {volume} {109}},\ \bibinfo {pages} {186404} (\bibinfo {year}
  {2012})}\BibitemShut {NoStop}%
\bibitem [{\citenamefont {Lan}\ \emph {et~al.}(2015)\citenamefont {Lan},
  \citenamefont {Kananenka},\ and\ \citenamefont {Zgid}}]{Lan2015}%
  \BibitemOpen
  \bibfield  {author} {\bibinfo {author} {\bibfnamefont {T.~N.}\ \bibnamefont
  {Lan}}, \bibinfo {author} {\bibfnamefont {A.~A.}\ \bibnamefont {Kananenka}},
  \ and\ \bibinfo {author} {\bibfnamefont {D.}~\bibnamefont {Zgid}},\ }\href
  {\doibase 10.1063/1.4938562} {\bibfield  {journal} {\bibinfo  {journal} {The
  Journal of Chemical Physics}\ }\textbf {\bibinfo {volume} {143}},\ \bibinfo
  {pages} {241102} (\bibinfo {year} {2015})},\ \Eprint
  {http://arxiv.org/abs/https://doi.org/10.1063/1.4938562}
  {https://doi.org/10.1063/1.4938562} \BibitemShut {NoStop}%
\bibitem [{\citenamefont {Bl{\"o}chl}(1994)}]{blochl1994projector}%
  \BibitemOpen
  \bibfield  {author} {\bibinfo {author} {\bibfnamefont {P.~E.}\ \bibnamefont
  {Bl{\"o}chl}},\ }\href@noop {} {\bibfield  {journal} {\bibinfo  {journal}
  {Physical review B}\ }\textbf {\bibinfo {volume} {50}},\ \bibinfo {pages}
  {17953} (\bibinfo {year} {1994})}\BibitemShut {NoStop}%
\bibitem [{\citenamefont {Kresse}\ and\ \citenamefont
  {Joubert}(1999)}]{kresse1999ultrasoft}%
  \BibitemOpen
  \bibfield  {author} {\bibinfo {author} {\bibfnamefont {G.}~\bibnamefont
  {Kresse}}\ and\ \bibinfo {author} {\bibfnamefont {D.}~\bibnamefont
  {Joubert}},\ }\href@noop {} {\bibfield  {journal} {\bibinfo  {journal}
  {Physical Review B}\ }\textbf {\bibinfo {volume} {59}},\ \bibinfo {pages}
  {1758} (\bibinfo {year} {1999})}\BibitemShut {NoStop}%
\bibitem [{\citenamefont {Kresse}\ and\ \citenamefont
  {Furthm{\"u}ller}(1996{\natexlab{a}})}]{kresse1996efficiency}%
  \BibitemOpen
  \bibfield  {author} {\bibinfo {author} {\bibfnamefont {G.}~\bibnamefont
  {Kresse}}\ and\ \bibinfo {author} {\bibfnamefont {J.}~\bibnamefont
  {Furthm{\"u}ller}},\ }\href@noop {} {\bibfield  {journal} {\bibinfo
  {journal} {Computational materials science}\ }\textbf {\bibinfo {volume}
  {6}},\ \bibinfo {pages} {15} (\bibinfo {year}
  {1996}{\natexlab{a}})}\BibitemShut {NoStop}%
\bibitem [{\citenamefont {Kresse}\ and\ \citenamefont
  {Furthm{\"u}ller}(1996{\natexlab{b}})}]{kresse1996efficient}%
  \BibitemOpen
  \bibfield  {author} {\bibinfo {author} {\bibfnamefont {G.}~\bibnamefont
  {Kresse}}\ and\ \bibinfo {author} {\bibfnamefont {J.}~\bibnamefont
  {Furthm{\"u}ller}},\ }\href@noop {} {\bibfield  {journal} {\bibinfo
  {journal} {Physical review B}\ }\textbf {\bibinfo {volume} {54}},\ \bibinfo
  {pages} {11169} (\bibinfo {year} {1996}{\natexlab{b}})}\BibitemShut {NoStop}%
\bibitem [{\citenamefont {Ruiz}\ \emph {et~al.}(1999)\citenamefont {Ruiz},
  \citenamefont {Cano}, \citenamefont {Alvarez},\ and\ \citenamefont
  {Alemany}}]{ruiz1999broken}%
  \BibitemOpen
  \bibfield  {author} {\bibinfo {author} {\bibfnamefont {E.}~\bibnamefont
  {Ruiz}}, \bibinfo {author} {\bibfnamefont {J.}~\bibnamefont {Cano}}, \bibinfo
  {author} {\bibfnamefont {S.}~\bibnamefont {Alvarez}}, \ and\ \bibinfo
  {author} {\bibfnamefont {P.}~\bibnamefont {Alemany}},\ }\href@noop {}
  {\bibfield  {journal} {\bibinfo  {journal} {Journal of computational
  chemistry}\ }\textbf {\bibinfo {volume} {20}},\ \bibinfo {pages} {1391}
  (\bibinfo {year} {1999})}\BibitemShut {NoStop}%
\bibitem [{\citenamefont {Marzari}\ \emph {et~al.}(2012)\citenamefont
  {Marzari}, \citenamefont {Mostofi}, \citenamefont {Yates}, \citenamefont
  {Souza},\ and\ \citenamefont {Vanderbilt}}]{marzari2012maximally}%
  \BibitemOpen
  \bibfield  {author} {\bibinfo {author} {\bibfnamefont {N.}~\bibnamefont
  {Marzari}}, \bibinfo {author} {\bibfnamefont {A.~A.}\ \bibnamefont
  {Mostofi}}, \bibinfo {author} {\bibfnamefont {J.~R.}\ \bibnamefont {Yates}},
  \bibinfo {author} {\bibfnamefont {I.}~\bibnamefont {Souza}}, \ and\ \bibinfo
  {author} {\bibfnamefont {D.}~\bibnamefont {Vanderbilt}},\ }\href@noop {}
  {\bibfield  {journal} {\bibinfo  {journal} {Reviews of Modern Physics}\
  }\textbf {\bibinfo {volume} {84}},\ \bibinfo {pages} {1419} (\bibinfo {year}
  {2012})}\BibitemShut {NoStop}%
\bibitem [{\citenamefont {Marzari}\ and\ \citenamefont
  {Vanderbilt}(1997)}]{marzari1997maximally}%
  \BibitemOpen
  \bibfield  {author} {\bibinfo {author} {\bibfnamefont {N.}~\bibnamefont
  {Marzari}}\ and\ \bibinfo {author} {\bibfnamefont {D.}~\bibnamefont
  {Vanderbilt}},\ }\href@noop {} {\bibfield  {journal} {\bibinfo  {journal}
  {Physical review B}\ }\textbf {\bibinfo {volume} {56}},\ \bibinfo {pages}
  {12847} (\bibinfo {year} {1997})}\BibitemShut {NoStop}%
\bibitem [{\citenamefont {Kanamori}(1963)}]{Kanamori1963}%
  \BibitemOpen
  \bibfield  {author} {\bibinfo {author} {\bibfnamefont {J.}~\bibnamefont
  {Kanamori}},\ }\href {\doibase 10.1143/PTP.30.275} {\bibfield  {journal}
  {\bibinfo  {journal} {Progress of Theoretical Physics}\ }\textbf {\bibinfo
  {volume} {30}},\ \bibinfo {pages} {275} (\bibinfo {year} {1963})}\BibitemShut
  {NoStop}%
\bibitem [{\citenamefont {Anisimov}\ \emph {et~al.}(1993)\citenamefont
  {Anisimov}, \citenamefont {Solovyev}, \citenamefont {Korotin}, \citenamefont
  {Czy\ifmmode~\dot{z}\else \.{z}\fi{}yk},\ and\ \citenamefont
  {Sawatzky}}]{Anisimov1993}%
  \BibitemOpen
  \bibfield  {author} {\bibinfo {author} {\bibfnamefont {V.~I.}\ \bibnamefont
  {Anisimov}}, \bibinfo {author} {\bibfnamefont {I.~V.}\ \bibnamefont
  {Solovyev}}, \bibinfo {author} {\bibfnamefont {M.~A.}\ \bibnamefont
  {Korotin}}, \bibinfo {author} {\bibfnamefont {M.~T.}\ \bibnamefont
  {Czy\ifmmode~\dot{z}\else \.{z}\fi{}yk}}, \ and\ \bibinfo {author}
  {\bibfnamefont {G.~A.}\ \bibnamefont {Sawatzky}},\ }\href {\doibase
  10.1103/PhysRevB.48.16929} {\bibfield  {journal} {\bibinfo  {journal} {Phys.
  Rev. B}\ }\textbf {\bibinfo {volume} {48}},\ \bibinfo {pages} {16929}
  (\bibinfo {year} {1993})}\BibitemShut {NoStop}%
\bibitem [{\citenamefont {{David Sherrill}}\ and\ \citenamefont
  {Schaefer}(1999)}]{DAVIDSHERRILL1999}%
  \BibitemOpen
  \bibfield  {author} {\bibinfo {author} {\bibfnamefont {C.}~\bibnamefont
  {{David Sherrill}}}\ and\ \bibinfo {author} {\bibfnamefont {H.~F.}\
  \bibnamefont {Schaefer}}\ }(\bibinfo  {publisher} {Academic Press},\ \bibinfo
  {year} {1999})\ pp.\ \bibinfo {pages} {143 -- 269}\BibitemShut {NoStop}%
\bibitem [{\citenamefont {Mostofi}\ \emph {et~al.}(2014)\citenamefont
  {Mostofi}, \citenamefont {Yates}, \citenamefont {Pizzi}, \citenamefont {Lee},
  \citenamefont {Souza}, \citenamefont {Vanderbilt},\ and\ \citenamefont
  {Marzari}}]{Mostofi2014}%
  \BibitemOpen
  \bibfield  {author} {\bibinfo {author} {\bibfnamefont {A.~A.}\ \bibnamefont
  {Mostofi}}, \bibinfo {author} {\bibfnamefont {J.~R.}\ \bibnamefont {Yates}},
  \bibinfo {author} {\bibfnamefont {G.}~\bibnamefont {Pizzi}}, \bibinfo
  {author} {\bibfnamefont {Y.-S.}\ \bibnamefont {Lee}}, \bibinfo {author}
  {\bibfnamefont {I.}~\bibnamefont {Souza}}, \bibinfo {author} {\bibfnamefont
  {D.}~\bibnamefont {Vanderbilt}}, \ and\ \bibinfo {author} {\bibfnamefont
  {N.}~\bibnamefont {Marzari}},\ }\href {\doibase
  https://doi.org/10.1016/j.cpc.2014.05.003} {\bibfield  {journal} {\bibinfo
  {journal} {Computer Physics Communications}\ }\textbf {\bibinfo {volume}
  {185}},\ \bibinfo {pages} {2309 } (\bibinfo {year} {2014})}\BibitemShut
  {NoStop}%
\end{thebibliography}%


\begin{thebibliography}{6}
\expandafter\ifx\csname natexlab\endcsname\relax\def\natexlab#1{#1}\fi
\expandafter\ifx\csname bibnamefont\endcsname\relax
  \def\bibnamefont#1{#1}\fi
\expandafter\ifx\csname bibfnamefont\endcsname\relax
  \def\bibfnamefont#1{#1}\fi
\expandafter\ifx\csname citenamefont\endcsname\relax
  \def\citenamefont#1{#1}\fi
\expandafter\ifx\csname url\endcsname\relax
  \def\url#1{\texttt{#1}}\fi
\expandafter\ifx\csname urlprefix\endcsname\relax\def\urlprefix{URL }\fi
\providecommand{\bibinfo}[2]{#2}
\providecommand{\eprint}[2][]{\url{#2}}

\bibitem[{\citenamefont{Das~Gupta et~al.}(2020)\citenamefont{Das~Gupta,
  Stewart, Chen, Abboud, Cheng, Hill, and Christou}}]{DasGupta2020}
\bibinfo{author}{\bibfnamefont{S.}~\bibnamefont{Das~Gupta}},
  \bibinfo{author}{\bibfnamefont{R.~L.} \bibnamefont{Stewart}},
  \bibinfo{author}{\bibfnamefont{D.-T.} \bibnamefont{Chen}},
  \bibinfo{author}{\bibfnamefont{K.~A.} \bibnamefont{Abboud}},
  \bibinfo{author}{\bibfnamefont{H.-P.} \bibnamefont{Cheng}},
  \bibinfo{author}{\bibfnamefont{S.}~\bibnamefont{Hill}}, \bibnamefont{and}
  \bibinfo{author}{\bibfnamefont{G.}~\bibnamefont{Christou}},
  \bibinfo{journal}{Inorganic Chemistry} \textbf{\bibinfo{volume}{59}},
  \bibinfo{pages}{8716} (\bibinfo{year}{2020}), ISSN \bibinfo{issn}{0020-1669},
  \urlprefix\url{https://doi.org/10.1021/acs.inorgchem.0c00332}.

\bibitem[{\citenamefont{Tang et~al.}(2009)\citenamefont{Tang, Sanville, and
  Henkelman}}]{tang2009grid}
\bibinfo{author}{\bibfnamefont{W.}~\bibnamefont{Tang}},
  \bibinfo{author}{\bibfnamefont{E.}~\bibnamefont{Sanville}}, \bibnamefont{and}
  \bibinfo{author}{\bibfnamefont{G.}~\bibnamefont{Henkelman}},
  \bibinfo{journal}{Journal of Physics: Condensed Matter}
  \textbf{\bibinfo{volume}{21}}, \bibinfo{pages}{084204}
  (\bibinfo{year}{2009}).

\bibitem[{\citenamefont{Handy}(1980)}]{Handy1980}
\bibinfo{author}{\bibfnamefont{N.}~\bibnamefont{Handy}},
  \bibinfo{journal}{Chemical Physics Letters} \textbf{\bibinfo{volume}{74}},
  \bibinfo{pages}{280 } (\bibinfo{year}{1980}), ISSN \bibinfo{issn}{0009-2614},
  \urlprefix\url{http://www.sciencedirect.com/science/article/pii/000926148085158X}.

\bibitem[{\citenamefont{Dudarev et~al.}(1998)\citenamefont{Dudarev, Botton,
  Savrasov, Humphreys, and Sutton}}]{Dudarev1998}
\bibinfo{author}{\bibfnamefont{S.~L.} \bibnamefont{Dudarev}},
  \bibinfo{author}{\bibfnamefont{G.~A.} \bibnamefont{Botton}},
  \bibinfo{author}{\bibfnamefont{S.~Y.} \bibnamefont{Savrasov}},
  \bibinfo{author}{\bibfnamefont{C.~J.} \bibnamefont{Humphreys}},
  \bibnamefont{and} \bibinfo{author}{\bibfnamefont{A.~P.}
  \bibnamefont{Sutton}}, \bibinfo{journal}{Phys. Rev. B}
  \textbf{\bibinfo{volume}{57}}, \bibinfo{pages}{1505} (\bibinfo{year}{1998}),
  \urlprefix\url{https://link.aps.org/doi/10.1103/PhysRevB.57.1505}.

\bibitem[{\citenamefont{Virtanen et~al.}(2020)\citenamefont{Virtanen, Gommers,
  Oliphant, Haberland, Reddy, Cournapeau, Burovski, Peterson, Weckesser, Bright
  et~al.}}]{2020SciPy-NMeth}
\bibinfo{author}{\bibfnamefont{P.}~\bibnamefont{Virtanen}},
  \bibinfo{author}{\bibfnamefont{R.}~\bibnamefont{Gommers}},
  \bibinfo{author}{\bibfnamefont{T.~E.} \bibnamefont{Oliphant}},
  \bibinfo{author}{\bibfnamefont{M.}~\bibnamefont{Haberland}},
  \bibinfo{author}{\bibfnamefont{T.}~\bibnamefont{Reddy}},
  \bibinfo{author}{\bibfnamefont{D.}~\bibnamefont{Cournapeau}},
  \bibinfo{author}{\bibfnamefont{E.}~\bibnamefont{Burovski}},
  \bibinfo{author}{\bibfnamefont{P.}~\bibnamefont{Peterson}},
  \bibinfo{author}{\bibfnamefont{W.}~\bibnamefont{Weckesser}},
  \bibinfo{author}{\bibfnamefont{J.}~\bibnamefont{Bright}},
  \bibnamefont{et~al.}, \bibinfo{journal}{Nature Methods}
  \textbf{\bibinfo{volume}{17}}, \bibinfo{pages}{261} (\bibinfo{year}{2020}).

\bibitem[{\citenamefont{Alexandropoulos
  et~al.}(2014)\citenamefont{Alexandropoulos, Mowson, Pilkington, Bekiari,
  Christou, and Stamatatos}}]{alexandropoulos2014emissive}
\bibinfo{author}{\bibfnamefont{D.~I.} \bibnamefont{Alexandropoulos}},
  \bibinfo{author}{\bibfnamefont{A.~M.} \bibnamefont{Mowson}},
  \bibinfo{author}{\bibfnamefont{M.}~\bibnamefont{Pilkington}},
  \bibinfo{author}{\bibfnamefont{V.}~\bibnamefont{Bekiari}},
  \bibinfo{author}{\bibfnamefont{G.}~\bibnamefont{Christou}}, \bibnamefont{and}
  \bibinfo{author}{\bibfnamefont{T.~C.} \bibnamefont{Stamatatos}},
  \bibinfo{journal}{Dalton Transactions} \textbf{\bibinfo{volume}{43}},
  \bibinfo{pages}{1965} (\bibinfo{year}{2014}).

\end{thebibliography}
\end{document}


\beginsupplement
\title{Supplementary Information}
\author{Dian-Teng Chen}
\affiliation{Department of Physics and the Quantum Theory Project, University of Florida, Gainesville, Florida 32611, USA}

\author{Jia Chen}
\affiliation{Department of Physics and the Quantum Theory Project, University of Florida, Gainesville, Florida 32611, USA}

\author{Xiang-Guo Li}
\affiliation{Department of Physics and the Quantum Theory Project, University of Florida, Gainesville, Florida 32611, USA}

\author{George Christou}
\affiliation{Department of Chemistry, University of Florida, Gainesville, Florida 32611, USA}

\author{Xiao-Guang Zhang}
\affiliation{Department of Physics and the Quantum Theory Project, University of Florida, Gainesville, Florida 32611, USA}

\author{Hai-Ping Cheng}
 \email{hping@ufl.edu}
\affiliation{Department of Physics and the Quantum Theory Project, University of Florida, Gainesville, Florida 32611, USA}

\maketitle

\section{Energies from DFT calculations}
The DFT energies for different spin configurations are listed in Table \ref{table:Mn5Ce3_1_configs}. DFT+$U$ calculations ($U_\Mn = 4.0 \eV$ and $U_\Mn = 4.4 \eV$) show that the ground state of \ch{Mn5Ce3} is a FM state, with a total spin of $S=17/2$. 
\begin{table}[hbt!]
\centering
\begin{tabular}{|p{15mm}|p{35mm}|p{35mm}|p{35mm}|p{15mm}|} 
 \hline
 Spin order (12345)& $E$/meV \newline ($U_\Mn = 4.0 \eV$ \newline $U_\Ce = 4.0 \eV$) & $E$/meV\newline ($U_\Mn = 4.4 \eV$ \newline $ U_\Ce = 2.0 \eV$ ) & $E$/meV \newline ($U_\Mn = 0 \eV $\newline $ U_\Ce = 2.0 \eV$)   & $M/\mu_B$  \\
 \hline
 $uuuuu$  &0 & 0 &  0       & 17 \\
 \hline
 $ddduu$  & 6 & 6 &  3       & $-5$ \\
\hline
 $dduuu$  &86  & 90 & 29       & 1 \\
 \hline
 $duddu$  & 75   & 79. & 25      & $-3$ \\
 \hline
 $duduu$  & 40    & 43 &  6      & 3 \\
 \hline
$duudu$  & 66 & 70 &  2      & 3 \\
\hline
 $duuud$  & 68  & 73 &  4     & 3 \\
 \hline
 $udddd$  & 31 & 35 &$-16$      & $-9$ \\
 \hline
 $udduu$  & 30 & 33 &$-18$      & 3 \\
 \hline
 $ududu$  & 74 & 78  & 23     & 3 \\
\hline
 $uduuu$   & 36 & 38  &  2     & 9 \\
 \hline
 $uuddu$  & 121 & 126   & 50    & 5 \\
 \hline
 $uudud$  & 120 & 125  & 48    & 5 \\
 \hline
 $uuduu$  & 83 & 86  & 28     & 11 \\
\hline
 $uuudu$  & 38 & 40   & 20     & 11 \\
 \hline
 $uuuud$  & 40   & 41  & 22    & 11 \\ 
\hline
\end{tabular}
\caption{Total energies $E$ and magnetization $M$ for each spin configuration of \ch{Mn5Ce3}. ``$u$'' denotes spin up and ``$d$'' spin down. The energy of the FM state is set to be zero. Energies for $4.4 \eV $ and $U_\Mn = 0 \eV$ in the third and fourth columns have been published in the supplementary document of the previous paper on \ch{Mn5Ce3} \cite{DasGupta2020}. They are included to make the table more complete.}
\label{table:Mn5Ce3_1_configs}
\end{table}


\section{Exchange coupling constants reduction}

\begin{figure}[hbt!]
\includegraphics[width=0.3\columnwidth]{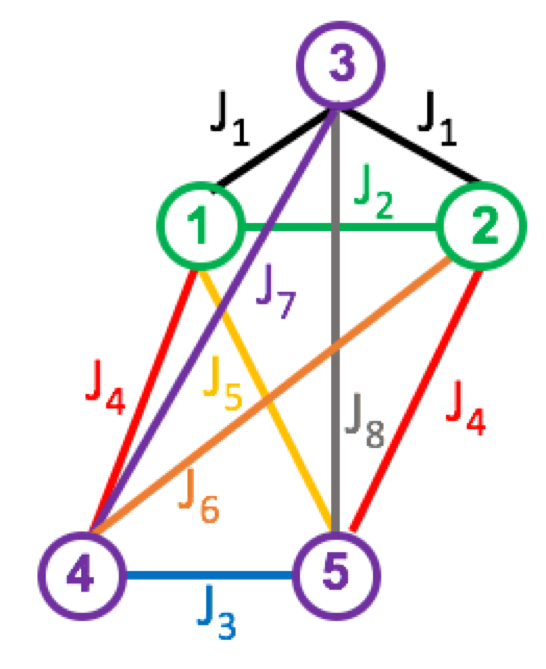}
\caption{The eight-$J$'s schematic diagram of exchange interactions between the Mn atoms in \ch{Mn5Ce3}. Different line colors represent different $J$'s. Distances between Mn atoms (by the order of $J_1$--$J_8$) are: $d_{23}$ = 2.94 \AA, $d_{13}$ = 3.02 \AA, $d_{12}$ =  4.15 \AA, $d_{45}$ = 2.74 \AA, $d_{25}$ = 6.06 \AA, $d_{14}$ = 5.89 \AA, $d_{15}$ = 5.99 \AA, $d_{24}$ = 7.00 \AA, $d_{34}$ = 8.04 \AA, $d_{35}$ = 7.22 \AA. The ground state is a FM state with $S=17/2$.}
\label{fig:Mn5Ce3_1_8j}
\end{figure}

To extract all the possible exchange interactions, we start with using eight different coupling constants in \ch{Mn5Ce3} as shown in Fig.~\ref{fig:Mn5Ce3_1_8j}. In principle, there can be a maximum of ten different exchange coupling constants for a molecule with five magnetic centers. As Mn1-Mn3 and Mn2-Mn3 pairs are both a  \ch{Mn^3+}-\ch{Mn^4+} pair and their Mn-Mn bond lengths are very close, we consider them both as $J_1$, and we consider both Mn1-Mn4 and Mn2-Mn5 pairs as $J_4$ for the same reason. As the results for these exchange couplings in Table \ref{table:Mn5Ce3_1_8j} show, $J_4$, $J_5$ and $J_6$ are similar, weak FM interactions, we can consider them as the same $J$. Similarly, we can also consider $J_7$ and $J_8$ as essentially the same type of $J$, as they both have the longest distances and are relatively weak. Thus, we further reduce the number of $J$'s from eight to five.

\begin{table}[hbt!]
\centering
\begin{tabular}{|p{15mm}|p{15mm}|p{25mm}|p{25mm}|p{25mm}|} 
 \hline
$ J$ types & Expt & $U_\Mn = 4.0 \eV $\newline $U_\Ce = 4.0 \eV$& $U_\Mn = 4.4 \eV $\newline $U_\Ce = 2.0 \eV$ & $U_\Mn = 0 \eV $\newline $U_\Ce = 2.0 \eV $  \\
 \hline
 $J_1$  & 34.4 & 43.8 & 45.7 & 14.8       \\
 \hline
 $J_2$  & $-12.8$ & $-7.5$ & $-6.5$  & $-17.3$     \\
\hline
 $J_3$  & 42.0 & 48.7 & 50.6  & 26.9      \\
\hline
 $J_4$  & $-0.4$ & 1.3 & 1.6   & 0.5        \\
\hline
 $J_5$  & 2.5  &1.9 & 2.0  & 1.3         \\
\hline
 $J_6$  & 0.0 & 0.3  & 0.1    & $-0.3$      \\
\hline
 $J_7$  & $-0.2$ & 0.0  & $-0.2$  & $-0.7$       \\
\hline
 $J_8$  & 3.0 & 2.0  & 1.8  & 2.0        \\
\hline
\end{tabular}
\caption{The fitted exchange coupling constants $J$ ($\cm^{-1}$) of \ch{Mn5Ce3} from the first-principles total energy calculations and from the experimental susceptibility curve, using eight different $J$'s as in  Fig.~\ref{fig:Mn5Ce3_1_8j}. Positive values mean ferromagnetic coupling.}
\label{table:Mn5Ce3_1_8j}
\end{table}

\clearpage

\section{Bader charge analysis}

A Bader charge analysis\cite{tang2009grid} is performed on \ch{Mn5Ce3} to calculate the charge distribution on the Manganese and Cerium ions.

\begin{table}[hbt!]
\centering
\begin{tabular}{|p{10mm}|p{12mm}|p{12mm}|p{12mm}|p{12mm}|p{12mm}|p{12mm}|p{12mm}|p{12mm}|} 
 \hline
  & \multicolumn{2}{p{24mm}|}{$U_\Mn = 4.0 \eV$\newline $U_\Ce = 4.0 \eV$}  & \multicolumn{2}{p{24mm}|}{$U_\Mn = 4.4\eV$\newline $U_\Ce = 2.0 \eV$}& \multicolumn{2}{p{24mm}|}{$U_\Mn = 0 \eV$\newline $U_\Ce = 2.0 \eV$} & \multicolumn{2}{p{24mm}|}{B3LYP}  \\
 \hline
  Atom & Charge &$M/\mu_B$ & Charge &$M/\mu_B$ &Charge &$M/\mu_B$& Charge &$M/\mu_B$  \\
 \hline
 Mn1  &5.30&3.84 & 5.30&3.87 & 5.36&3.52     & 5.25&3.68  \\
 \hline
 Mn2  & 5.31&3.85 & 5.31&3.88 & 5.36&3.54   & 5.25&3.70  \\
\hline
 Mn3  &5.22&3.07 & 5.21&3.11 & 5.27&2.71     & 5.14&2.88   \\
\hline
 Mn4  & 5.23&3.05 &  5.23&3.09 & 5.29&2.67     & 5.16&2.86    \\
\hline
 Mn5  & 5.22&3.04  & 5.22&3.08  & 5.28&2.66     & 5.15&2.85    \\
 \hline
 Ce1  & 9.52&0.05 & 9.57&0.06 & 9.57&0.11  & 9.42&0.04    \\
 \hline
 Ce2  & 9.54&0.05 & 9.59&0.05  & 9.60&0.10     & 9.44 &0.04   \\
 \hline
 Ce3  & 9.53&0.06 & 9.58&0.07 & 9.59&0.14      & 9.44 &0.05   \\
\hline
\end{tabular}
\caption{Bader charge and partial magnetic moment on Mn and Ce ions of \ch{Mn5Ce3}.}
\label{table:Bader}
\end{table}


\section{Selected Configuration Interaction Calculation}

In configuration interaction calculations carried out in this work, the wavefunction is expressed as linear combination of Slater determinants: ($\hat{S}_z$ eigenfunctions) 

\begin{align}\label{CIWF}
|\Psi\rangle_{\textrm{CI}} = \sum_{I} C_{I} |\Phi_{\textrm{I}}\rangle
\end{align}

Each determinant is expressed as alpha and beta strings \cite{Handy1980}
\begin{align}\label{determinant}
|\Phi_{\textrm{I}}\rangle = c^{\dagger}_{1\uparrow} c^{\dagger}_{2\uparrow}\cdots c^{\dagger}_{1\downarrow} c^{\dagger}_{2\downarrow} \cdots|0\rangle = | \alpha(1,2,\cdots) \beta(1,2,\cdots) \rangle
\end{align}
For example, the reference was constructed as:
\begin{align}\label{ref}
|\Phi_{\textrm{ref}}\rangle = c^{\dagger}_{\textrm{Mn}_{1}\uparrow} \cdots c^{\dagger}_{\textrm{Mn}_{17}\uparrow} c^{\dagger}_{\textrm{O}_{1}\uparrow} \cdots c^{\dagger}_{\textrm{O}_{35}\uparrow}c^{\dagger}_{\textrm{O}_{36}\uparrow} c^{\dagger}_{\textrm{O}_{1}\downarrow} \cdots c^{\dagger}_{\textrm{O}_{35}\downarrow} c^{\dagger}_{\textrm{O}_{36}\downarrow}|0\rangle
\end{align}
with all Mn, Ce and Oxygen ions in charge states listed as $[\Mn_2^{\,3+} \Mn_3^{\,4+} \Ce_3^{\,4+} \O_9^{\,2-}$]; and spin states for Mn ions are symmetry breaking and fully polarized. 
The Hamiltonian matrix element are evaluated as:
\begin{align}\label{Hmatrix}
H_{IJ} = \langle \Phi_I | \hat{H} | \Phi_J \rangle
\end{align}
Where $\hat{H}$ is defined as Eq.~4 in the paper. 
When the on-site interaction and double-counting potentials are only density-density interactions, which is the situation in this work, the evaluation of off-diagonal elements can be simplified, 
\begin{align}\label{off-diag}
H_{I\neq J} = \langle \Phi_I | \hat{H}_{\textrm{hop}} | \Phi_J \rangle,
\end{align}
because density operators do not change the determinant. In this work, $U=5.0$ eV and $J= 1.0$ eV were used for both Mn and Ce, and the effective $U$ is 4.0 eV.\cite{Dudarev1998} A phase factor (a sign in this case) may arise when applying the hopping operator, which needs to be tracked. The Hamiltonian matrix is diagonalized by the  \textbf{SciPy} library, \cite{2020SciPy-NMeth} and the ground state vector from diagonalization gives us the coefficient for each determinant in Eq.~\ref{CIWF}

We did a test of the DFT+selected CI method in the single-molecule magnet $[\Mn^{\,3+}_3 \O(\O_2\CMe)_3 (\mpko)_3](\Cl\O_4)$ \cite{alexandropoulos2014emissive}. In this molecule, all three Mn ions have the same charge state. In the results, superexchange is considered as the mechanism. Coefficients for excitations can be found in Fig.~\ref{Mn3}. Similar to the \ch{Mn5Ce3} molecule, charge transfers between Mn ions are less significant than those between $\O^{2-}$ and $\Mn^{3+}$ ions, which supports the conclusion that superexchange is the more important exchange mechanism.

\begin{figure}\label{Mn3}

\includegraphics[width=\columnwidth]{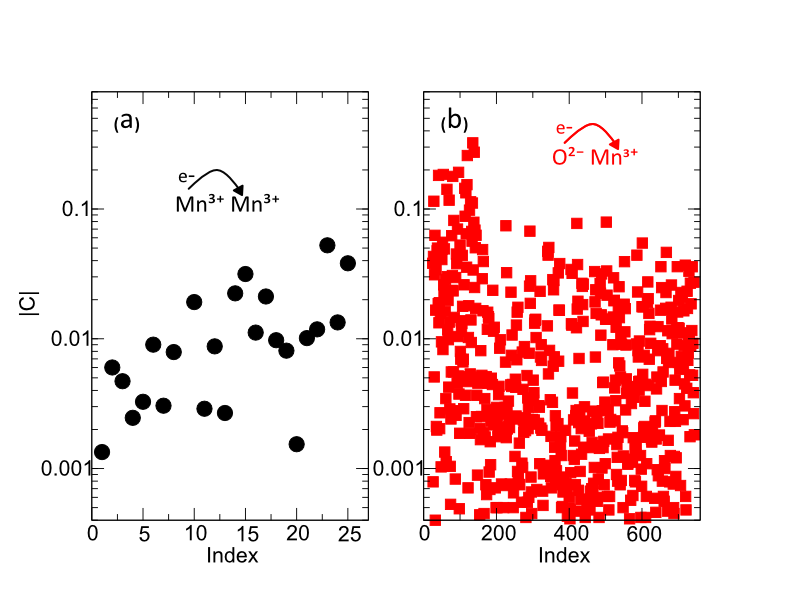}
\caption{Absolute value of coefficients for excitations obtained from diagonalization plotted on a logarithmic scale. 
Panel (a): Coefficients for excitations that transfer electrons between $\Mn^{3+}$. 
Panel (b): Coefficients for excitations that transfer electrons from $\O^{2-}$ to $\Mn^{3+}$. }
\label{Mn3}
\end{figure}

\bibliography{supplement}